\documentclass[pdflatex,sn-mathphys]{sn-jnl}

\jyear{2021}%

\theoremstyle{thmstyleone}%
\theoremstyle{thmstyletwo}%

\theoremstyle{thmstylethree}%

\usepackage{multirow}
\begin{document}

\title[The Simons Observatory 220 and 280\,GHz Focal-Plane Module]{The Simons Observatory 220 and 280\,GHz Focal-Plane Module: Design and Initial Characterization}


\author*[1]{\fnm{Erin} \sur{Healy}}\email{ehealy@princeton.edu}
\author[1]{\fnm{Daniel} \sur{Dutcher}}
\author[1]{\fnm{Zachary} \sur{Atkins}}
\author[2]{\fnm{Jason} \sur{Austermann}}
\author[3,4]{\fnm{Steve K.} \sur{Choi}}
\author[3]{\fnm{Cody J.} \sur{Duell}}
\author[2]{\fnm{Shannon} \sur{Duff}}
\author[5]{\fnm{Nicholas} \sur{Galitzki}}
\author[3]{\fnm{Zachary B.} \sur{Huber}}
\author[2]{\fnm{Johannes} \sur{Hubmayr}}
\author[6]{\fnm{Bradley R.} \sur{Johnson}}
\author[1]{\fnm{Heather} \sur{McCarrick}}
\author[3,4,7]{\fnm{Michael D.} \sur{Niemack}}
\author[1]{\fnm{Rita} \sur{Sonka}}
\author[1]{\fnm{Suzanne T.} \sur{Staggs}}
\author[3]{\fnm{Eve} \sur{Vavagiakis}}
\author[1]{\fnm{Yuhan} \sur{Wang}}
\author[8]{\fnm{Zhilei} \sur{Xu}}
\author[1]{\fnm{Kaiwen} \sur{Zheng}}

\affil[1]{Joseph Henry Laboratories, Jadwin Hall, Princeton University, Princeton, NJ, USA}
\affil[2]{National Institute of Standards and Technology, Boulder, CO, USA}
\affil[3]{Department of Physics, Cornell University, Ithaca, NY, USA}
\affil[4]{Department of Astronomy, Cornell University, Ithaca, NY, USA}
\affil[5]{Department of Physics, University of California San Diego, La Jolla, CA, USA}
\affil[6]{University of Virginia, Department of Astronomy, Charlottesville, VA, USA}
\affil[7]{Kavli Institute at Cornell for Nanoscale Science, Cornell University, Ithaca, NY, USA}
\affil[8]{MIT Kavli Institute, Massachusetts Institute of Technology, Cambridge, MA, USA}


\abstract{The Simons Observatory (SO) will detect and map the temperature and polarization of the millimeter-wavelength sky from Cerro Toco, Chile across a range of angular scales, providing rich data sets for cosmological and astrophysical analysis. The SO focal planes will be tiled with compact hexagonal packages, called Universal Focal-plane Modules (UFMs), in which the transition-edge sensor (TES) detectors are coupled to 100\,mK microwave-multiplexing electronics. Three different types of dichroic TES detector arrays with bands centered at 30/40, 90/150, and 220/280\,GHz will be implemented across the 49 planned UFMs. The 90/150\,GHz and 220/280\,GHz arrays each contain 1,764 TESes, which are read out with two 910x multiplexer circuits. The modules contain a series of densely routed silicon chips, which are packaged together in a controlled electromagnetic environment with robust heat-sinking to 100\,mK. Following an overview of the module design, we report on early results from the first 220/280\,GHz UFM, including detector yield, as well as readout and detector noise levels.}




\maketitle

\section{Introduction}\label{sec1}
The millimeter-wavelength sky contains rich information about the origins and evolution of our universe. The temperature and polarization patterns in the cosmic microwave background (CMB) give us a precise snapshot of this cosmological epoch, about 380,000 years after the big bang. Simons Observatory (SO)\footnote{https://simonsobservatory.org/} will build four telescopes, one large-aperture telescope (LAT)~\cite{Zhu_2021} and three small-aperture telescopes (SATs)~\cite{Galitzki_2018}, to observe the millimeter sky from the Atacama Desert in Chile at 5,200\,m. The telescopes will observe at six frequencies between 30 and 280\,GHz with over 60,000 transition-edge sensor (TES) detectors. This paper focuses on the highest frequency detector modules, which have dichroic pixels centered at 220 and 280\,GHz.  

The small-aperture survey aims to constrain the tensor-to-scalar ratio \emph{r}, which parameterizes the primordial gravity-wave polarization signal~\cite{SO}. Measuring in the 220/280\,GHz band enables the removal of foreground sources, such as galactic dust, in CMB polarization maps. The large-aperture survey will make small-angular-scale measurements of the CMB temperature anisotropies and polarization~\cite{SO}. One target of this survey is to use the Sunyaev-Zeldovich effects to study structure in the later universe. The thermal-SZ effect, in which CMB photons interact with high-temperature electrons, has a frequency-dependence such that 220 and 280\,GHz measurements will measure the null and peak of this spectrum.

The Simons Observatory focal-plane modules are called Universal Focal-plane Modules (UFMs) for their universality between the two types of telescope receivers (LAT and SAT). The UFMs are hexagonal packages that couple the transition-edge sensor detectors to the telescope optics and the 100\,mK microwave-multiplexing readout components. There will be three types of UFMs corresponding to the SO frequency bands: a low frequency (LF) array with pixels centered at 30/40\,GHz, a mid frequency (MF) array at 90/150\,GHz, and a ultra-high frequency (UHF)\footnote{For historical reasons, SO calls the highest detector frequency band the ``ultra-high" frequency to distinguish it from Atacama Cosmology Telescope (ACT) high frequency (HF) arrays.} array at 220/280\,GHz. The common multiplexer design and overall architecture has been published [4]. The MF modules are reported on in a concurrent publication \cite{mfpaper}. Thirteen UHF focal-plane modules will be built for use in both a SAT (7x UFMs) and LAT (6x UFMs).

\section{Methods}\label{UFMs}

\subsection{Detectors and optical coupling}\label{Detectors}
The UHF detector array wafer, which is fabricated at NIST out of a 150\,mm silicon wafer, consists of 432 pixels, each of which is sensitive to two frequencies (220 and 280\,GHz) and two orthogonal polarizations. At the center of each pixel is the polarization sensor: the orthomode transducer (OMT), which couples orthogonal linear polarizations of incident photons to superconducting wiring lithographed on the detector wafer. These signals are then filtered by frequency to one of the four TESes in the pixel. Each detector array, thereby, has a total of 1,728 optical TESes, split evenly among the two frequencies. There are an additional 36 dark detectors, which aid in detector calibration and are not coupled to OMTs. 

A monolithic gold-plated aluminum feed-horn array with spline-profiled horns efficiently couples incident power to the OMTs while preserving beam symmetry~\cite{Simon}. The horn array also has the flange that interfaces to the focal plane in the optics tube, therefore setting the focus of the beam. There are four additional wafers that are part of the optical coupling stack that complete the waveguide from the feed horn to the OMT and provide a quarter-wavelength reflection to improve coupling efficiency.


\subsection{Readout components}\label{Readout components}
 The SO readout technology, microwave multiplexing, works by coupling each detector to a microwave resonator with a unique frequency between 4 and 6\,GHz via a superconducting quantum interference device (SQUID). These tones are coupled to a common transmission line. The signals are amplified and demodulated with SLAC microresonator radio frequency (SMuRF) electronics~\cite{Henderson18}. With this architecture, up to 910 detector channels can be read out with a single pair of coaxial cables. Because the UHF detector wafers house 1,764 detectors, two 910x multiplexing circuits are used to read out a single UFM (some readout channels are left uncoupled to detectors).
 
Central to the SO multiplexing architecture are the NIST microwave-multiplexer ($\mu$mux or mux) chips, which contain the 100\,mK readout components, including the microwave resonators, SQUIDs, and coupling inductors~\cite{McCarrick}. The multiplexer chips, which each have 65 channels, are fabricated in sets of 14 chips with complementary resonators spanning the 4 to 6\,GHz readout bandwidth. These chips are packaged in the Universal Microwave-mulitplexing Module (UMM), which sits behind the detector stack in the UFM. A silicon routing wafer distributes the electrical signals between the mux chips, the detector wafer, and circuits that connect to optics tube wiring. The mux chips and routing wafer are packaged in a copper box that reduces box-mode coupling and provides robust grounding. 

\begin{figure}[ht]
\centering
\includegraphics[width=0.72\textwidth]{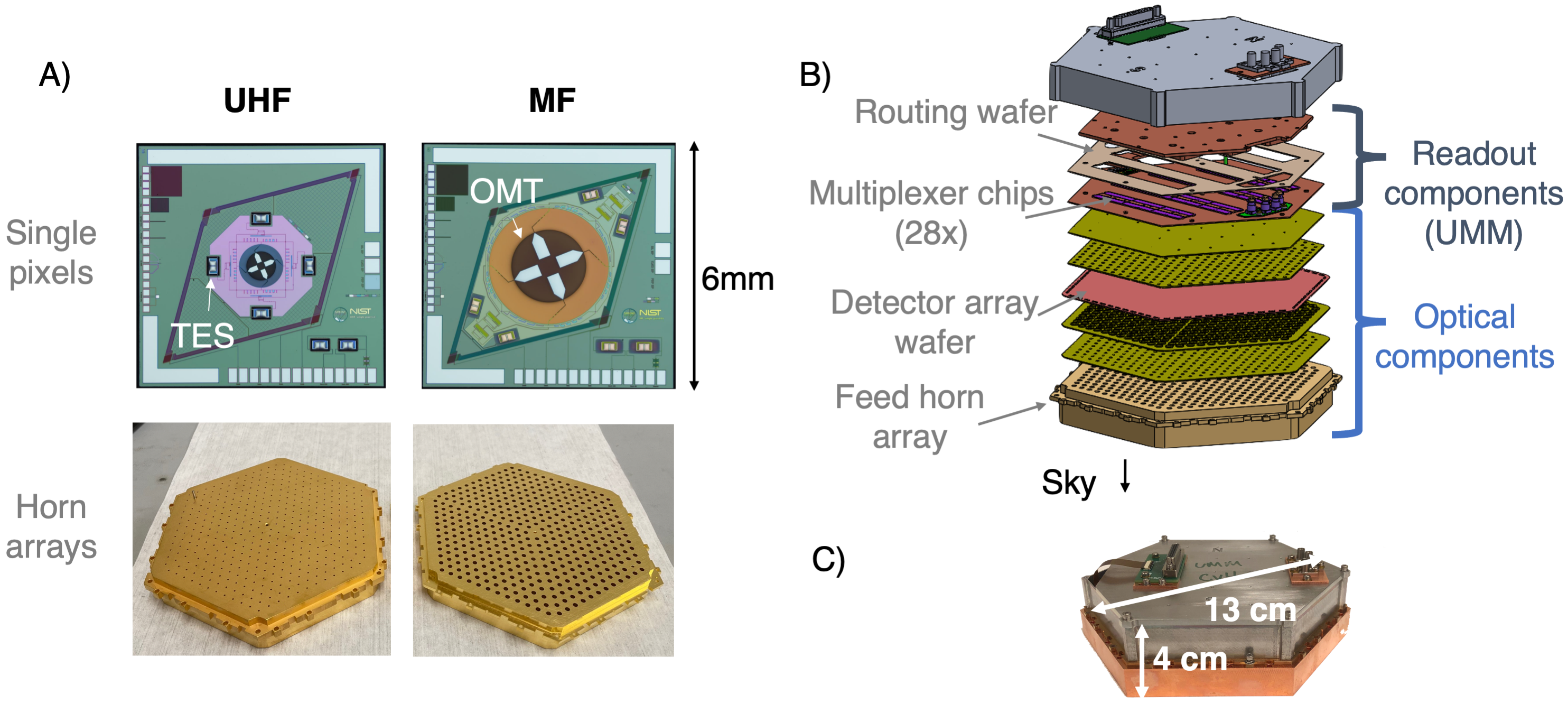}
\caption{A: Comparison of the UHF and MF detectors and horn arrays. Top: Single pixel chips used for detector development. The circumscribed cross is the orthomode transducer (OMT), which is surrounded by four TESes. Bottom: The UHF and MF horn arrays are shown, as viewed from the detector wafer side. B: Exploded view of the SO focal plane modules, which package the microwave-multiplexing readout components behind the optical stack. C: A fully assembled UFM.}\label{fig1}
\end{figure}

\subsection{UHF and MF UFMs}\label{MF vs. UHF}

The UHF UFM design closely mirrors the MF UFM design, as they both use the same readout and optical coupling technologies. On the readout side, the UMMs are completely interchangeable between the MF and UHF UFMs. The optical coupling, however, depends on the wavelength, such that the UHF has a smaller antenna (OMT) size, different horn profile, and different flange height to align the detectors to the telescope focus. A comparison of the MF and UHF detectors and horn arrays can be found in Fig. \ref{fig1}.

\subsection{The first UHF module}\label{Assembly}
\begin{figure}[h]%
\centering
\includegraphics[width=0.7\textwidth]{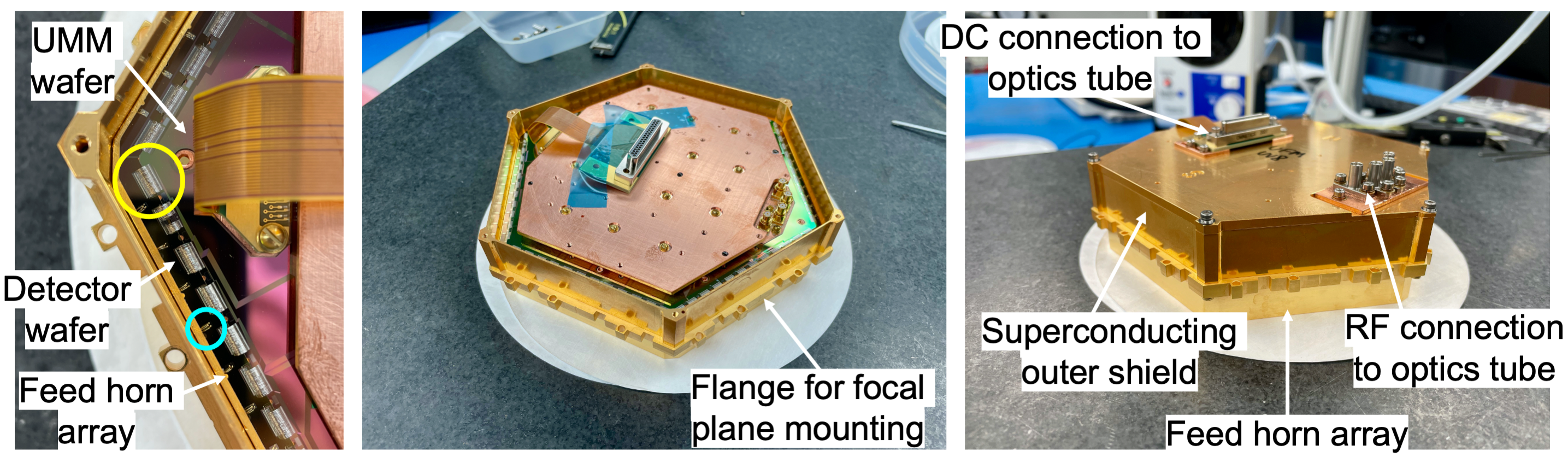}
\caption{Photos from UFM-Uv8 assembly. Left: Photo of UFM. The aluminum wire bonds between multiplexer (routing wafer) and detector wafer that connect the TESes to the readout circuit are circled in yellow. Gold bonds between the detector wafer and gold-plated feed horn array are circled in cyan. Both the gold bonds and aluminum TES bonds surround the perimeter. Center: The UFM before the gold-plated aluminum shield is added. Right: The UFM after all of the assembly has been completed. }\label{fig2}
\end{figure}

The first prototype module, labeled UFM-Uv8, was assembled from a tested UMM and the first SO UHF detector wafer fabricated at NIST. The UMM was assembled according to standardized protocols~\cite{Healy}, which were developed to minimize variation between the modules. Integrating the UHF detector optical stack involved additional assembly, including precision alignment of the detectors to the feed horn array with dowel pins, adding gold wire bonds between the detector wafer and horn array for heat sinking, and electrically coupling the UMM with aluminum wire bonds around the perimeter. Photos from the UFM-Uv8 assembly can be seen in Fig. \ref{fig2}.

\section{Results}\label{testing}
\begin{figure}[ht]%
\centering
\includegraphics[width=0.75\textwidth]{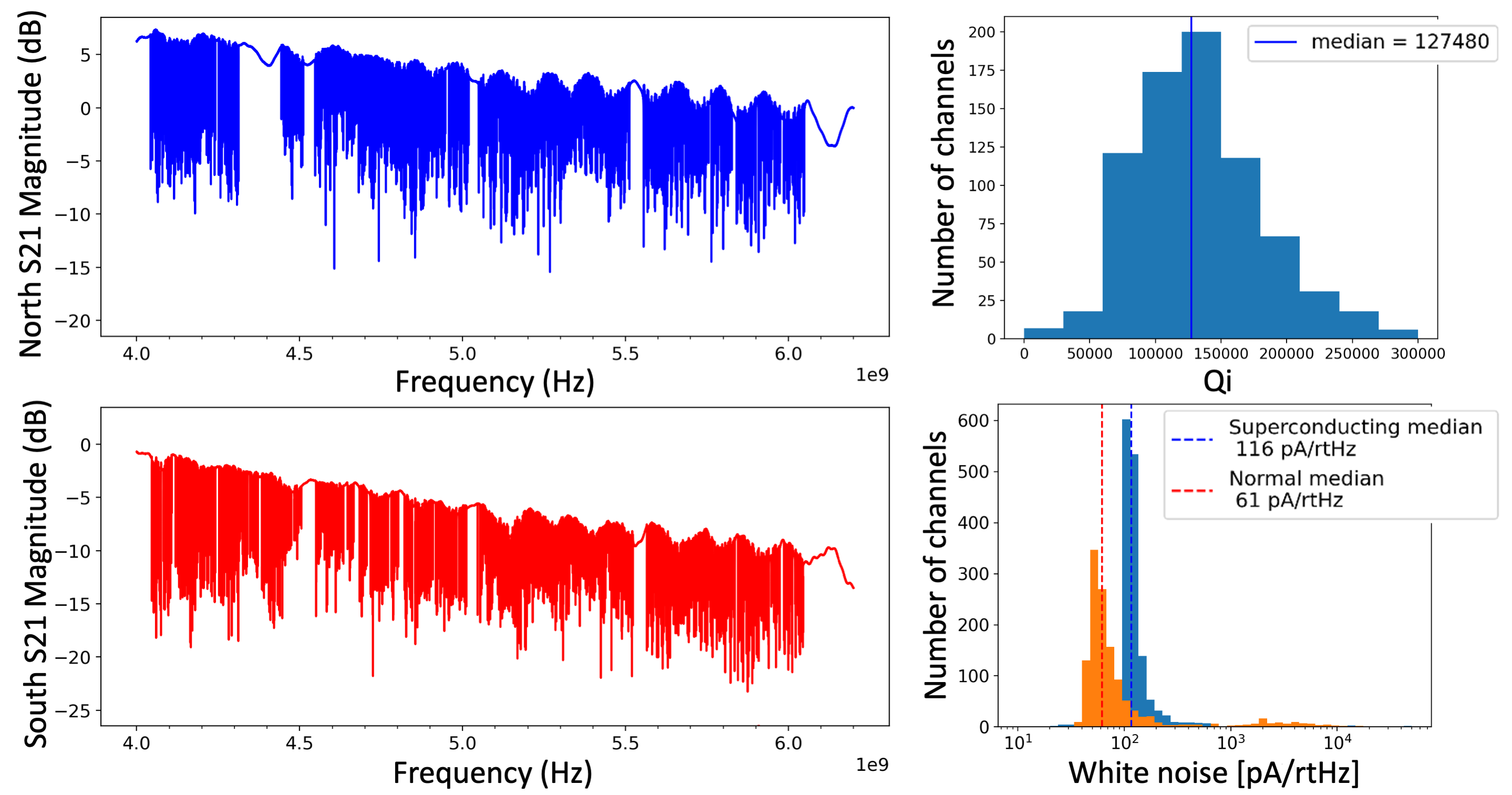}
\caption{Left: transmission (S21) plots for the North and South multiplexers of the prototype UHF UFM. Each dip in the transmission corresponds to a resonator, i.e. a readout channel. Upper right: Histogram of resonator internal quality factor ($Q_i$). Lower right: Histogram of channel noise when the detectors are superconducting (blue) and normal (orange). The noise data were measured with SMuRF. The datasets analysed during the study are available from the corresponding author on reasonable request.}\label{fig3}
\end{figure}

The prototype UHF module, UFM-Uv8, was tested in a dilution refrigerator equipped with a microwave SQUID multiplexing readout chain similar to the SO design~\cite{Rao}. The UFM has two readout chains, which are labeled ``North" and ``South," each of which are coupled to half of the 1,764 TES channels. The UFM was first measured with a vector network analyzer (VNA) to determine the S-parameters as a function of frequency, which can be seen in Figure \ref{fig3}. These plots validate the RF environment of the multiplexer, as we see a relatively flat transmission and deep resonances (median dip depth -8\,dB) at frequencies that correspond to a resonator channel. Histograms of the per-channel internal quality factor ($Q_i$) can also be found in Figure \ref{fig3}. Table \ref{tab1} reports the resonator yield statistics.

Following the VNA measurement, the UFM was tested with SMuRF electronics. These warm electronics provide the bias power for the detectors, flux ramp modulation for the SQUIDs, and GHz probe tones for the resonators. We used SMuRF to make noise measurements when the detectors were superconducting and normal. Plots of the noise can be found in Figure \ref{fig3}. We then took I-V curves on the detectors, i.e. measured the current through the TES while voltage-ramping through the superconducting transition, to confirm their operability. We found that the transition temperature for the detectors was 160-170\,mK, within the SO specification range. Table \ref{tab2} shows the detector-readout coupling and detector response yield.

\begin{table}[h]
\caption{Resonator yield statistics for the prototype UHF module. Nominally, each half of the UFM will have 910 resonators when all 28 multiplexer chips are present. Note that one of the chips in UFM-Uv8  was damaged, which accounts for the 4\% loss in the final assembly. The final column reports the number of resonator readout channels that were identified with SMuRF. }\label{tab1}
\begin{tabular}{|p{2.1cm}|p{2.1cm}|p{2.1cm}|p{2.6cm}|}
 \hline
 \multicolumn{4}{|c|}{Resonator yield} \\
 \hline
 \emph{Module half} & \emph{Nominal number} & \emph{Number in final assembly} & \emph{Number measured with SMuRF} \\
 \hline
 North   & 910  & 846 & 791 \\
 South   & 910  & 910 & 841 \\
 \hline
 Total   & 1820  & 1756 & 1632 \\
  Percentage & & \textbf{96\%} & \textbf{90\%}\\
 \hline
\end{tabular}
\end{table}

\begin{table}[h]
\caption{Detector yield statistics for prototype UHF module. The number of wired channels corresponds to how many detectors were nominally coupled to resonators. To determine which the detectors were in fact coupled to the readout, a sine wave was input on the bias line and the corresponding resonator channel read out with SMuRF. The number of I-V curves indicates the number of detectors that responded to voltage ramping through their transition. Note that due to an understood issue with a prototype SMuRF component, one bias group with 160 detectors was not operable, which accounts for most of the 10\% yield drop between the sine-wave and I-V-curve response.}\label{tab2}
\begin{tabular}{|p{2.2cm}|p{2.2cm}|p{2.3cm}|p{2.2cm}|}
 \hline
  \multicolumn{4}{|c|}{Detector yield} \\
  \hline
  \emph{Detector frequency}  & \emph{Number of wired channels} & \emph{Number of sine wave responses} & \emph{Number of IV curves} \\
  \hline
   220\,GHz  & 900  & 793 & 774 \\
   280\,GHz  & 800  & 681 & 534 \\
   \hline
   Total   & 1700  & 1474 & 1308 \\
  Percentage   & \textbf{96\%}  & \textbf{84\%} & \textbf{74\%} \\
 \hline
\end{tabular}
\end{table}

\section{Discussion}\label{sec12}
With this first screening of UFM-Uv8, we have confirmed some of the fundamental features of the design: the multiplexer coupling to detectors was validated, detectors were able to be operated, and the overall noise and yield numbers meet the SO specifications. We plan to further characterize the optical performance of this array, including measuring detector response to a cold blackbody source to determine saturation powers and optical efficiency. The prototype array will also be used to validate the optics of the SO UHF receivers.

\backmatter


\bmhead{Acknowledgements}
This work was supported in part by a grant from the Simons Foundation (Award \#457687, B.K.). ZBH acknowledge support from the NASA Space Technology Graduate Research Opportunities Award. SKC acknowledges support from NSF award AST-2001866. Zhilei Xu is supported by the Gordon and Betty Moore Foundation through grant GBMF5215 to the Massachusetts Institute of Technology.


\bibliography{LTDbib}


\end{document}